\documentclass[12pt]{article}
\usepackage[utf8]{inputenc}
 
\usepackage{rotating}
\usepackage{booktabs}
\usepackage{natbib}
\usepackage{graphicx}
\usepackage{booktabs}
\usepackage{setspace}

\usepackage{pa}

\usepackage[fleqn]{amsmath}               
\usepackage{dsfont}

\usepackage{float}
\usepackage{tabularx}

\usepackage[titletoc,title]{appendix}
\usepackage[labelfont=bf]{caption}

\usepackage[margin=0.98in]{geometry}
\usepackage{enumitem}

\usepackage[hang,flushmargin]{footmisc}

\usepackage{amsfonts}

\begin{document}


\title{Choosing Imputation Models}

\author{Moritz Marbach\thanks{Texas A\&M University, The Bush School of Government \& Public Service,  moritz.marbach@tamu.edu} }
\date{}

\maketitle

\begin{abstract}
Imputing missing values is an important preprocessing step in data analysis, but the literature offers little guidance on how to choose between different imputation models. This letter suggests adopting the imputation model that generates a density of imputed values most similar to those of the observed values for an incomplete variable after balancing all other covariates. We recommend stable balancing weights as a practical approach to balance covariates whose distribution is expected to differ if the values are not missing completely at random. After balancing, discrepancy statistics can be used to compare the density of imputed and observed values. We illustrate the application of the suggested approach using simulated and real-world survey data from the American National Election Study, comparing popular imputation approaches including random forests, hot-deck, predictive mean matching, and multivariate normal imputation. An R package implementing the suggested approach accompanies this letter. 
\end{abstract}

\newpage


Missing data are ubiquitous, and missing data imputation remains an important preprocessing step in any data analysis. A large number of different multiple imputation models have been developed over the years. Popular examples for univariate multiple imputation methods include parametric regression imputation \citep{Rubin.1987}, as well as non-parametric approaches such as predictive mean matching imputation \citep{Little.1988b}, random hot-deck imputation \citep{Andridge.Little.2010, Cranmer.Gill.2013} and random forest imputation \citep{Doove.et.al.2014, Stekhoven.Buehlmann.2012}. Multivariate multiple imputation approaches include multivariate normal imputation \citep{Schafer.1997,King.et.al.2001} as well as imputation using a sequence of chained univariate imputation methods \citep{vanBuuren.et.al.2006,vanBuuren.2007}.\footnote{Whereas multivariate normal imputation is implemented in the \texttt{Amelia} R package, chained equation models for imputation are available in the \texttt{MICE} package. In Stata both approaches are available through the command \texttt{mi impute mvn} and \texttt{mi impute chained}.} 

For users imputing a dataset, choosing among available imputation models is often difficult. As imputation is different from optimal point prediction \citep[e.g.,][]{Rubin.1996}, standard model choice approaches such as cross-validation, have limited use. In this letter, we point to a simple criterion to choose among different imputation approaches: choose the imputation model that produces imputed values that are more consistent with the assumed missing data mechanism. Specifically, for observations that only differ in that some values for one variable $y$ are missing, choose the imputation model that produces a density of imputed values that is most similar to the density of the observed values. To render this criterion operational, we suggest computing weights such that the densities of covariates $X$ for observations with and without missing values on $y$ are identical, and then utilizing discrepancy statistics to determine the differences between the density of imputed values and the (adjusted) density of observed values.

The concept of comparing the \emph{unadjusted} density of observed and imputed values is an established recommendation for identifying problems with imputation models \citep[e.g.,][]{Abayomi.et.al.2008,vanBuuren.2018}. However, an \emph{unadjusted} comparison often has limited use as it is expected that there will be differences in the densities if the missing data indicator is correlated with other covariates in the data, that is, if the data are not missing completely at random (MCAR) in the terminology of \citet{Little.Rubin.2019}. We therefore suggest weighting the data before making the comparison. 

Our approach is complementary to a recently proposed, graphical approach in the medical research literature which has not been used in social science research.\footnote{We reviewed all 33 citations to the paper and found no applied study in the fields of Political Science, Economics or Sociology.} Specifically, \citet{Bondarenko.Raghunathan.2016} propose plotting the imputed and observed values against the (estimated) probability that a value is missing and preferring imputations that display fewer differences in the density of observed and imputed values across various levels of the missingness probability. 

Given the limited popularity of this approach, we developed a general, simple---and, arguably, more intuitive---weighting method that allows users to choose an imputation model that suits their dataset best. More generally, we hope that the suggested approach facilitates the application of multiple imputation in applied research. In the following, we detail the assumptions needed, how to estimate the weights and illustrate the application using simulated and real-world data. An R package implementing the suggested approach accompanies this letter and is available on Github.\footnote{https://github.com/sumtxt/missDiag}

\section*{Adjusted Comparison of Imputed and Observed Values}

Consider a variable $Y$ for which some values are missing and some values are observed. A missing data indicator, $M$, encodes $M_i=0$ if a value $y_i$ is observed or $M_i=1$ if it is missing. Let $X$ be the set of all other covariates. We assume that the observations are independent and identically distributed (i.i.d.) and that the data are missing at random (MAR) \citep{Rubin.1976,Mealli.Rubin.2015}, that is, $f(M_i|x_i,y_i) = f(M_i|x_i) \forall i=1,..., N$. Most imputation models, including those mentioned in the introduction, rely on both these assumptions. 

When the data are i.i.d. and MAR, the conditional density of the missing values, $f(Y|X=x,M=0)$, is equal to the conditional density of the observed values, $f(Y|X=x,M=1)$ \citep[e.g.,][p. 18]{Little.Rubin.2019}. In other words, holding $X$ constant, the density of missing and observed values is equal under the MAR assumption. This notion suggests an assessment of whether the empirical densities of the \emph{imputed} values and observed values differ after adjusting for differences in $X$. Imputation models for which imputed values are more similar to the observed values after adjustment are more consistent with MAR and may be preferred over imputation models that lead to imputed values with larger deviations. 

To adjust for the covariates $X$, we propose computing a weight such that the moments of the covariate densities for observations with and without missing values in $Y$ match. Different weighting schemes have been proposed in the literature to compute such weights. For example, Hainmueller's entropy balancing weighting scheme computes such weights by minimizing the Kullback entropy divergence \citep{Hainmueller.2012} and Zubizarreta's weighting scheme minimizes the variance among the weights \citep{Zubizarreta.2015}. We prefer the latter as it, naturally, leads to fewer instances of extreme weights but note that entropy-balancing computations tend to be faster in practice.

Zubizarreta's weighting scheme, a convex quadratic programming problem that can be solved with standard optimization solvers, takes the following form: 

\begin{equation*}
\begin{split}
\underset{w}{\mathrm{minimize}} \quad 
	& \| w-\overline{w} \|_2^2 \\
\mathrm{subject\ to} \quad	 	
	& | w'x_{M=0,p} - \overline{x}_{M=1,p} | \leq \delta_p, \quad p=1,..., P, \\
	& \mathds{1}' w = 1,  \\
	& w \geq 0,
\end{split}
\end{equation*}

where $w$ is the vector of weights to be computed, $\overline{w}$ is the average weight, $\| \cdot \|_2$ is the Euclidean norm (the square root of the sum of squares), $x_{M=0,p}$ is the covariate vector $p$ of observations without a missing value on $y$, and $\overline{x}_{M=1,p}$ is the sample mean of covariate $p$ among observations for which $y$ is imputed. The tolerance parameter $\delta_p$ is typically set to a small value if not exactly 0. This weighting scheme balances the first moment of the covariate densities (the means). To balance higher-order moments one may include the appropriate terms. For example, to balance the second moment of $x_p$ (the variances) one includes $x_p^2$.

To identify the difference between the (weighted) density of imputed and observed values, we suggest adopting four statistics that are widely used to assess balance between the treatment and control groups in causal inference \citep{Imbens.Rubin.2015,Franklin.et.al.2014}. For all four metrics, smaller values suggest a higher similarity. 

The standardized mean difference (SMD) is the difference between the means of the densities for the imputed and observed values standardized by the square of their average variance. The log variance ratio, log(VR), is the logarithm of the variance ratio. The SMD measures the difference in the densities' locations, and the VR measures the difference in the densities' dispersion. 

The SMD and VR measure the differences in the first and second moments of two densities, whereas the Kolmogorov–-Smirnov (KS) statistic is also sensitive to deviations in higher moments. The KS measures the largest discrepancy between the empirical and cumulative distributions. It is best thought of as a measure of the largest dissimilarity between the imputed and observed values.

A complementary statistic is the overlap (OVL) coefficient. The OVL coefficient measures local similarities instead of local discrepancies between two densities. Specifically, it is defined as the proportion of the overlap between two densities. Typically, the difference 1-OVL is reported such that smaller values suggest a higher similarity.  

To properly reflect uncertainty about the imputed values in any analysis, a series of imputed datasets are typically generated; each dataset is analyzed and the estimates are combined using Rubin's rules \citep{Rubin.1987}. This procedure can also be applied to the discrepancy statistics introduced above. Users may average the discrepancy statistics across the multiple imputed datasets or, alternatively, compare the densities of the balance statistics graphically as illustrated below.

In practice, the covariates $X$ might also include missing values. Adopting a multivariate version of MAR \citep[cf.][p. 14]{Little.Rubin.2019}, one typically imputes missing values in $X$ jointly with missing values in $Y$. Therefore one can compute a weight such that the moments of the \emph{filled-in} covariate densities for observations with and without missing values in $Y$ match.\footnote{An alternative but limited strategy is to treat missingness as another category if $X$ includes categorical variables alone and compute the weight such that the moments of the binary indicators for each category (including the missing value category) for observations with and without missing values in $Y$ match.} One concern with this strategy is that the multivariate version of MAR implies that the joined density of missing values is equal to the joined density of observed values. Yet, applying the suggested approach amounts to comparing the (full) conditional densities of $Y$ alone rather than the joined densities. A more comprehensive assessment therefore involves the comparison of the conditional densities of $Y$ and all covariates $X$ one at a time. That said, it would be desirable for future research to develop an approach that compares the joined densities directly.

\section*{Illustration with Simulated Data}

In this section, we use simulated data to demonstrate how the above approach can be used to assess which imputation model produces better imputed data (i.e., generates a density of imputed values more similar to the [reweighted] density of observed values) and that pooled regression estimates from models fitted on better imputed data are also less biased. 

We simulate 1,000 datasets (with $N=1,000$) from a linear model of the form $y = x \cdot z + e$ with fixed parameters. Covariate $z$ is drawn from a Bernoulli distribution with mean 0.5 and continuous covariate $x$ from a uniform distribution with range -5 to 5. We set the variance of the normal error distribution to 1 and let the proportion of missing values in $y$ vary with $z$. The two proportions are drawn independently from a uniform distribution in the range of 0.1 to 0.5. The larger value determines the proportion of missing values if $z=1$ whereas the smaller value determines the proportion of missing values if $z=0$. The simulated missing data mechanism is consistent with the i.i.d. and MAR assumptions. 

Iterating over all simulated datasets, we impute the missing values five times using four popular imputation approaches implemented in the MICE software package \citep{vanBuuren.GroothuisOudshoorn.2011}: predictive mean matching, random hot-deck imputation, normal model imputation and random forest imputation. Predictive mean matching amounts to measuring the distance between all observed values and a missing value using their predicted values from a linear, additive regression model. Five observations with the smallest distance are used to construct a donor candidate pool, from which one observation is drawn at random to impute the missing value.

\begin{figure}[!ht]
\centering
\includegraphics[width=\textwidth]{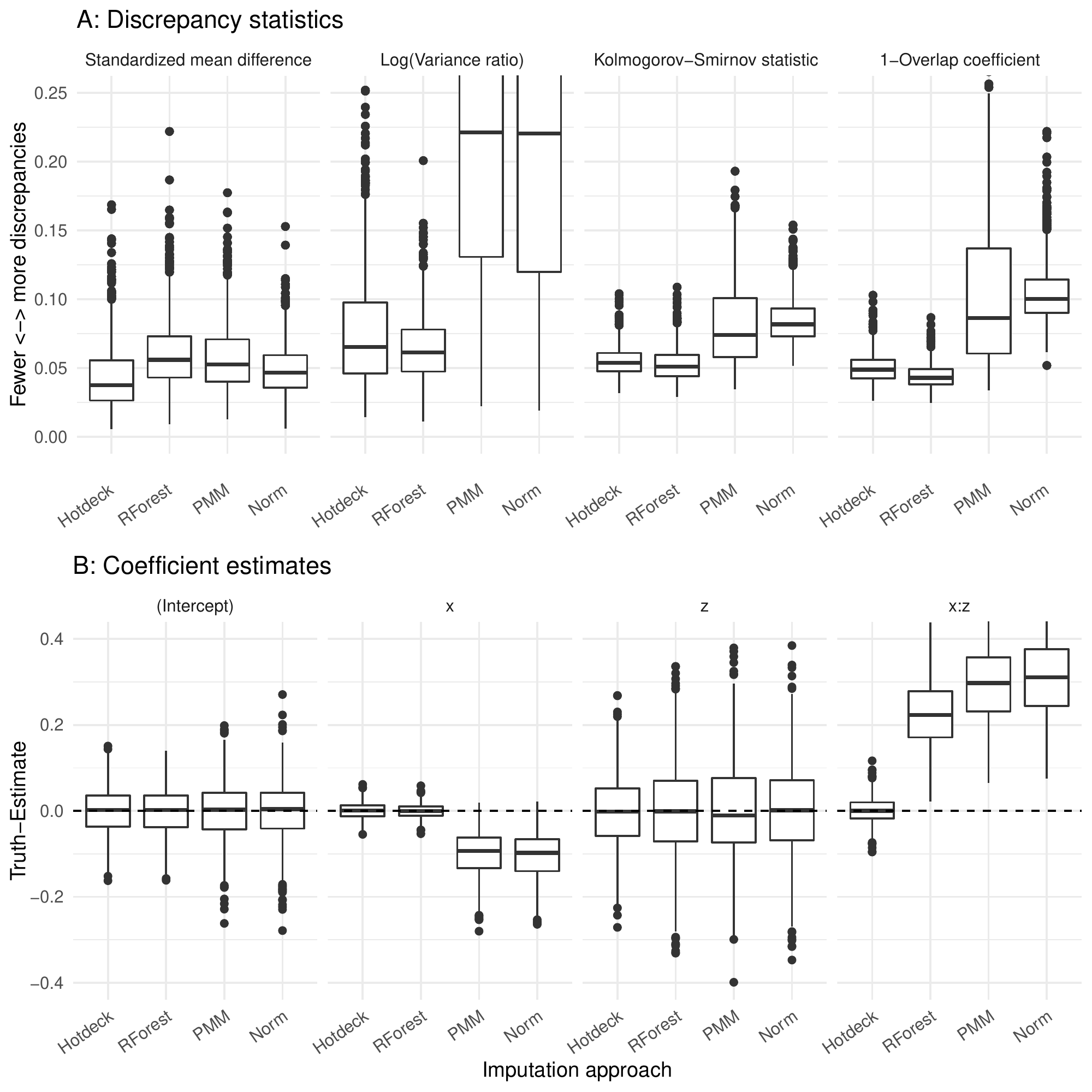}
\caption{Summary of 1,000 Monte Carlo simulations. Panel A: Box plots of four discrepancy statistics comparing the weighted density of observed and imputed values. Panel B: Box plots of the bias in the (pooled) coefficient estimates of a linear model $y = x \cdot z + \epsilon$. Each dataset was imputed five times using random hot-deck imputation (Hotdeck), random forest imputation (RForest), predictive mean matching (PMM) and normal model imputation (Norm). Some box plots are clipped to increase readability. \label{mc}}
\end{figure}

Random hot-deck imputation is similar to predictive mean matching but compares observations' covariate profile using a (multivariate) distance function such as the Mahalanobis distance. A random draw from the candidate pool of five observations is used to impute the missing value. Normal model imputation uses the predicted values from a linear, additive regression model to impute the missing values.

Random forest imputation is typically encountered as a single imputation method \citep[e.g.,][]{Stekhoven.Buehlmann.2012}, but an implementation for the multiple imputation context is also available \citep{Doove.et.al.2014}. Random forest is an ensemble machine learning method based on decision trees. Different from the normal model imputation, it does not rely on a linear, additive model and thus has the potential to accommodate more flexible dependencies between variables. 

We apply the diagnostic as described above and estimate the coefficients of a linear regression model that includes an interaction term. Consistent with the multiple imputation approach, we estimate these linear regression models on each imputed dataset and average the resulting coefficient estimates.

In Figure \ref{mc} (Panel A), each box plot describes a distribution of a discrepancy statistic (averaged across five imputed datasets). Given the non-linear data generating process, we would expect imputation approaches without linearity assumptions (hot-deck imputation and random forest imputation) to perform better. This is indeed what we find: The discrepancy statistics from these two imputation approaches are, on average, closer to zero, suggesting that hock-deck imputation and random forest imputation produces better imputed data for the simulated data generating process. 

The differences between random hot-deck imputation and random forest imputation are less pronounced. While the standardized mean differences from random hot-deck imputation are closer to zero on average, we see fewer differences for the other three discrepancy statistics. However, altogether the discrepancy statistics suggest that random hot-deck imputation produces better imputed data for the simulated data generating process.\footnote{The longer tailed distribution for the log(VR) distribution might cast doubt on the superior performance of random hot-deck imputation (relative to random forest imputation). However, random hot-deck imputation also does better in terms of the variance once we explicitly balance the second moment (the variance) of the covariate densities (see Figure \ref{mc_xx}).} 

Figure \ref{mc} (Panel B) shows the distribution of the bias (truth-estimate) in the (pooled) coefficient estimates for four linear regression terms across the simulations. As expected, coefficient estimates based on the better imputed data are, on average, less biased. When the data are imputed by random hot-deck there is, on average, no bias left in either the main effects or the interaction effect. When the data are imputed by the random forest, there is no bias left for the main effects on average, but some bias remains in the interaction effect. For the normal model imputation and predictive mean matching, the bias remains substantial in both the main effect and interaction effect.

\section*{Imputing ANES}

To illustrate the application of the suggested approach beyond simulated data, we use survey data from the 2008 edition of the American National Election Studies ($N=2,265$). The dataset prepared by \citet{Kropko.et.al.2014} includes 11 variables in different scales with various levels of missingness. Table \ref{tab} provides a description of the variables and the number of missing values. 

\begin{table}[ht!]
\begin{center}
\begin{tabularx}{\textwidth}{llX }
\toprule
 Variable   & Missing & Description \\ 
\midrule
Age   & 0 & Age of the respondent. \\ 
Sex   & 0 & Sex of the respondent. \\
Race  & 9 & Non-white vs. white. \\
Education  & 0 & No high school, some high school, high school diploma, college. \\
Income  & 158 & Low, medium, high. \\
Religion  & 393 & Protestant, Catholic/Orthodox, Atheist/other. \\
Marital status  & 0 & Single, married, no longer married. \\
Gov. support & 152 & 7 response categories ranging from ``Govt should let each person get ahead on own'' to ``Govt should see to jobs and standard of living.'' \\
Environment & 23 & Whether the respondent sees the environment as an important issue (yes/no). \\
Vote    & 258 & Vote choice in the 2008 Presidential election: Obama, McCain, No vote/Other. \\
Time  & 217 & Time for the respondent to complete the survey. \\ 
\bottomrule
\end{tabularx}
\caption{Summary of eleven variables from the 2008 American National Election Studies.\label{tab}}
\end{center}
\end{table}

We impute the dataset using the four most popular imputation approaches for mixed-type datasets again: predictive mean matching, random hot-deck imputation, random forest imputation and multivariate normal model imputation. We reply on the Amelia package \citep{Honaker.et.al.2011} for multivariate normal imputation and on the MICE package for the other three approaches \citep{vanBuuren.GroothuisOudshoorn.2011}. To simplify the analysis, we consider ``government support'' as a continuous variable. We also log-transform the variable ``response time'' before imputation.

\begin{figure}[!ht]
\centering
\includegraphics[width=\textwidth]{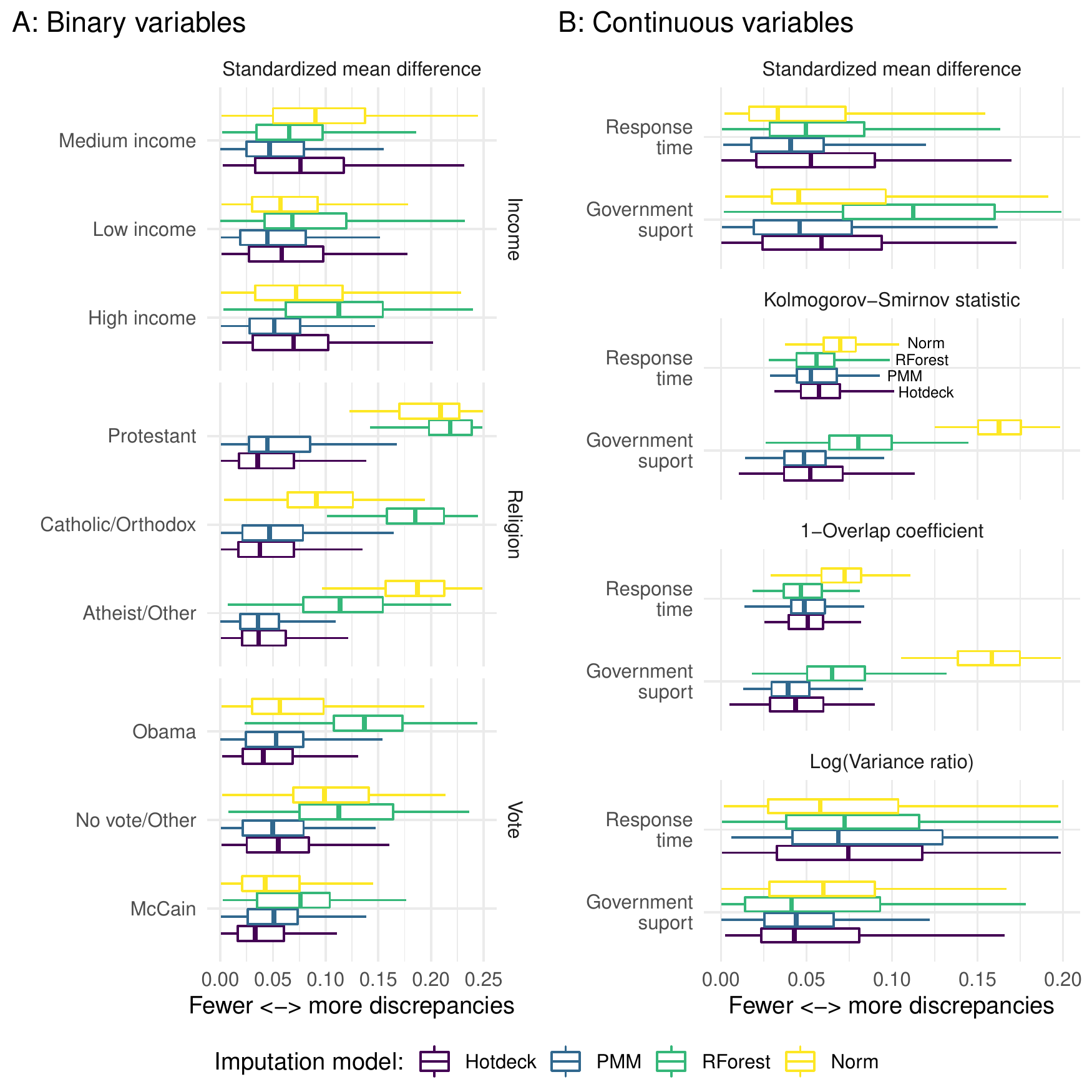}
\caption{Box plots of discrepancy statistics comparing the weighted density of observed and imputed values across 100 imputed datasets. Some outliers are removed, and some upper whiskers are clipped to increase readability. \label{fig}}
\end{figure}

For each variable with at least 25 imputed values, we construct weights that balance all other variables using the weighting scheme described previously. Figure \ref{fig} shows the density of the discrepancy statistics across 100 imputed datasets. As the mean of a binary variable is a sufficient statistic, we only provide the standardized mean difference for all binary variables.

We observe the largest discrepancies in the standardized mean differences for two of the three categories of the variable ``religion.'' The density of the standardized mean difference for the multivariate normal imputation and random forest are both bounded away from zero and are distinct from the density of the two other approaches. We observe similar yet less pronounced differences for the variable ``vote'' and ``government support.'' For two continuous variables we observe a similar pattern for the OVL coefficient and KS statistic; hot-deck and predictive mean matching imputation both do much better than the other two approaches. All four approaches are similar in terms of the variance ratio.

The hot-deck and predictive mean imputation performances are fairly similar across variables and discrepancy statistics. Table \ref{tab2} demonstrates that predictive mean matching does a little better than hot-deck imputation when averaging across imputed datasets and variables. Based on these averages one may recommend using predictive mean matching as it produces imputed values that are more consistent with the MAR assumption on average across all variables. Alternatively, one may use the results to revise the imputation model. MICE relies on chained univariate imputation models which means that it is easy to use different imputation approaches for different variables and we could therefore select predictive mean matching for some variables and hot-deck imputation for others, depending on their performance. 

\begin{table}[ht]
\centering
\begin{tabular}{lrrrr}
\toprule
& \multicolumn{4}{c}{Discrepancy statistics} \\ 
\cmidrule(lr){2-5}
& SMD & log(VR) & KS & 1-OVL \\ 
\midrule
PMM 		& 0.053 & 0.077 & 0.029 & 0.028 \\ 
Hotdeck & 0.058 & 0.085 & 0.031 & 0.030 \\ 
Norm 		& 0.100 & 0.082 & 0.061 & 0.061 \\ 
RForest & 0.125 & 0.079 & 0.059 & 0.057 \\ 
 \bottomrule
\end{tabular}
\caption{Average discrepancy statistics across variables and imputed datasets for predictive mean matching imputation (PMM), random hot-deck imputation (Hotdeck), multivariate normal imputation (Norm) and random forest imputation (RForest). The four displayed discrepancy statistics include standardized mean difference (SMD), variance ratio (VR), Kolmogorov–-Smirnov statistic (KS) and overlap coefficient (1-OVL). \label{tab2}}
\end{table}

The differential performance of the four imputation approaches cannot be explained by the differential performance of the constructed weights. In the appendix (Figure \ref{fig_sm_income}--\ref{fig_sm_vote}), we document that the constructed weights balance the first moment of the densities exactly (as expected). Some moderate imbalances are left for the higher moments of the continuous variables. In principle, these could be removed by adding, for example, the variables squared and cubed to the weighting scheme. However, because these higher-moment imbalances are largely not differential between the four imputation approaches, they cannot explain the differential performance.

One concern with the analysis might be that higher-order interactions are not explicitly balanced. For example, while the weights ensure that the proportions of Obama voters and of women among respondents with and without missing values on the variable religion are identical, they do not guarantee that the proportions of women voting for Obama are identical. To evaluate the sensitivity of the results to the inclusion of some higher-order interactions, we replicate the analysis balancing the interactions of the demographic variables (gender, age, and race) with all other variables. The results, which appear in Figure \ref{fig_int}, are largely identical to those in Figure \ref{fig}.

\section*{Conclusion}

Incomplete data is a common challenge when analyzing real-world data. Another challenge is choosing from many available multiple imputation approaches to fill-in the missing values. While some approaches are more popular than others, the literature offers little advice on how to choose among them for a given dataset. In the absence of such advice, list-wise deletion remains a popular default approach despite its known deficits. And even if multiple imputation is used in applied work, it is common practice to rely on the (currently) most popular approach in the literature \citep[cf.][]{Lall.2016} which may or may not be optimal for a given dataset. 

This letter provides a conceptually simple and practical approach to choosing between competing imputation models that impute data under MAR. The approach is flexible and does not have to be tailored to a particular imputation model; all it needs is the data with the missing values and the filled-in dataset. It builds on the same assumptions made by most multiple imputation models, in particular the MAR assumption. While more work is needed to understand how to choose between imputation models when the data are missing not at random (MNAR), we hope that this letter, in conjunction with the accompanying, easy-to-use software package in R, prompts users to make explicit comparisons between different imputation models for their dataset and to choose the one that imputes values most consistent with MAR. 

We also hope that this contribution lowers the barriers faced by applied researchers trying to address missing data in their data analysis and therefore reduces the reliance on list-wise deletion which always leads to inefficient estimates but typically also introduces bias. More generally, our approach highlights how weighting and imputation approaches, often seen as alternative strategies to handle missing data, can be combined in a productive manner. While we are not the first to combine both---for example, \citet{Seaman.et.al.2012} combine multiple imputations with weighting to achieve double robustness---the combination of weighting and imputation appears to be a fruitful avenue for future research.

\newpage

\addcontentsline{toc}{section}{References}
\bibliographystyle{chicago}
\bibliography{ref}

\begin{thebibliography}{}

\bibitem[\protect\citeauthoryear{Abayomi, Gelman, and Levy}{Abayomi
  et~al.}{2008}]{Abayomi.et.al.2008}
Abayomi, K., A.~Gelman, and M.~Levy (2008).
\newblock {Diagnostics for Multivariate Imputations}.
\newblock {\em Journal of the Royal Statistical Society. Series C\/}~{\em
  57\/}(3), 273--291.

\bibitem[\protect\citeauthoryear{Andridge and Little}{Andridge and
  Little}{2010}]{Andridge.Little.2010}
Andridge, R.~R. and R.~J. Little (2010).
\newblock {A Review Of Hot Deck Imputation For Survey Non-response}.
\newblock {\em International Statistical Review\/}~{\em 78\/}(1), 40--64.

\bibitem[\protect\citeauthoryear{Bondarenko and Raghunathan}{Bondarenko and
  Raghunathan}{2016}]{Bondarenko.Raghunathan.2016}
Bondarenko, I. and T.~Raghunathan (2016).
\newblock {Graphical and Numerical Diagnostic Tools to Assess Suitability of
  Multiple Imputations and Imputation Models}.
\newblock {\em Statistics in Medicine\/}~{\em 35\/}(17), 3007--3020.

\bibitem[\protect\citeauthoryear{Cranmer and Gill}{Cranmer and
  Gill}{2013}]{Cranmer.Gill.2013}
Cranmer, S.~J. and J.~Gill (2013).
\newblock {We Have to Be Discrete About This: A Non-Parametric Imputation
  Technique for Missing Categorical Data}.
\newblock {\em British Journal of Political Science\/}~{\em 43\/}(2), 425--449.

\bibitem[\protect\citeauthoryear{Doove, Van~Buuren, and Dusseldorp}{Doove
  et~al.}{2014}]{Doove.et.al.2014}
Doove, L.~L., S.~Van~Buuren, and E.~Dusseldorp (2014).
\newblock {Recursive Partitioning for Missing Data Imputation in the Presence
  of Interaction Effects}.
\newblock {\em {Computational Statistics \& Data Analysis}\/}~{\em 72},
  92--104.

\bibitem[\protect\citeauthoryear{Franklin, Rassen, Ackermann, Bartels, and
  Schneeweiss}{Franklin et~al.}{2014}]{Franklin.et.al.2014}
Franklin, J.~M., J.~A. Rassen, D.~Ackermann, D.~B. Bartels, and S.~Schneeweiss
  (2014).
\newblock {Metrics for Covariate Balance in Cohort Studies Of Causal Effects}.
\newblock {\em Statistics in Medicine\/}~{\em 33\/}(10), 1685--1699.

\bibitem[\protect\citeauthoryear{Hainmueller}{Hainmueller}{2012}]{Hainmueller.2012}
Hainmueller, J. (2012).
\newblock {Entropy Balancing for Causal Effects: A Multivariate Reweighting
  Method to Produce Balanced Samples In Observational Studies}.
\newblock {\em Political Analysis\/}~{\em 20\/}(1), 25--46.

\bibitem[\protect\citeauthoryear{Honaker, King, and Blackwell}{Honaker
  et~al.}{2011}]{Honaker.et.al.2011}
Honaker, J., G.~King, and M.~Blackwell (2011).
\newblock {{Amelia II}: A Program for Missing Data}.
\newblock {\em Journal of Statistical Software\/}~{\em 45\/}(7), 1--47.

\bibitem[\protect\citeauthoryear{Imbens and Rubin}{Imbens and
  Rubin}{2015}]{Imbens.Rubin.2015}
Imbens, G.~W. and D.~B. Rubin (2015).
\newblock {\em {Causal Inference in Statistics, Social, and Biomedical
  Sciences: An Introduction}}.
\newblock Cambridge: Cambridge University Press.

\bibitem[\protect\citeauthoryear{King, Honaker, Joseph, and Scheve}{King
  et~al.}{2001}]{King.et.al.2001}
King, G., J.~Honaker, A.~Joseph, and K.~Scheve (2001).
\newblock {Analyzing Incomplete Political Science Data: An Alternative
  Algorithm for Multiple Imputation}.
\newblock {\em American Political Science Review\/}~{\em 95\/}(1), 49--69.

\bibitem[\protect\citeauthoryear{Kropko, Goodrich, Gelman, and Hill}{Kropko
  et~al.}{2014}]{Kropko.et.al.2014}
Kropko, J., B.~Goodrich, A.~Gelman, and J.~Hill (2014).
\newblock {Multiple Imputation for Continuous and Categorical Data: Comparing
  Joint Multivariate Normal And Conditional Approaches}.
\newblock {\em Political Analysis\/}~{\em 22\/}(4), 497--519.

\bibitem[\protect\citeauthoryear{Lall}{Lall}{2016}]{Lall.2016}
Lall, R. (2016).
\newblock {How Multiple Imputation Makes a Difference}.
\newblock {\em Political Analysis\/}~{\em 24\/}(4), 414--433.

\bibitem[\protect\citeauthoryear{Little}{Little}{1988}]{Little.1988b}
Little, R.~J. (1988).
\newblock {Missing-data Adjustments in Large Surveys}.
\newblock {\em Journal of Business \& Economic Statistics\/}~{\em 6\/}(3),
  287--296.

\bibitem[\protect\citeauthoryear{Little and Rubin}{Little and
  Rubin}{2019}]{Little.Rubin.2019}
Little, R. J.~A. and D.~B. Rubin (2019).
\newblock {\em {Statistical Analysis with Missing Data}\/} (Third Edition ed.).
\newblock New York: Wiley.

\bibitem[\protect\citeauthoryear{Mealli and Rubin}{Mealli and
  Rubin}{2015}]{Mealli.Rubin.2015}
Mealli, F. and D.~B. Rubin (2015).
\newblock {Clarifying Missing at Random and Related Definitions, and
  Implications When Coupled With Exchangeability}.
\newblock {\em Biometrika\/}~{\em 102\/}(4), 995--1000.

\bibitem[\protect\citeauthoryear{Rubin}{Rubin}{1976}]{Rubin.1976}
Rubin, D.~B. (1976).
\newblock {Inference and Missing Data}.
\newblock {\em Biometrika\/}~{\em 63\/}(3), 581--592.

\bibitem[\protect\citeauthoryear{Rubin}{Rubin}{1987}]{Rubin.1987}
Rubin, D.~B. (1987).
\newblock {\em {Multiple Imputation for Nonresponse in Surveys}}.
\newblock New York: J. Wiley \& Sons.

\bibitem[\protect\citeauthoryear{Rubin}{Rubin}{1996}]{Rubin.1996}
Rubin, D.~B. (1996).
\newblock {Multiple Imputation After 18+ Years}.
\newblock {\em Journal of the American Statistical Association\/}~{\em
  91\/}(434), 473--489.

\bibitem[\protect\citeauthoryear{Schafer}{Schafer}{1997}]{Schafer.1997}
Schafer, J.~L. (1997).
\newblock {\em {Analysis of Incomplete Multivariate Data}}.
\newblock Boca Raton: Chapman and Hall.

\bibitem[\protect\citeauthoryear{Seaman, White, Copas, and Li}{Seaman
  et~al.}{2012}]{Seaman.et.al.2012}
Seaman, S.~R., I.~R. White, A.~J. Copas, and L.~Li (2012).
\newblock {Combining Multiple Imputation and Inverse-Probability Weighting}.
\newblock {\em Biometrics\/}~{\em 68\/}(1), 129--137.

\bibitem[\protect\citeauthoryear{Stekhoven and B{\"u}hlmann}{Stekhoven and
  B{\"u}hlmann}{2012}]{Stekhoven.Buehlmann.2012}
Stekhoven, D.~J. and P.~B{\"u}hlmann (2012).
\newblock {MissForest---Non-parametric Missing Value Imputation for Mixed-Type
  Data}.
\newblock {\em Bioinformatics\/}~{\em 28\/}(1), 112--118.

\bibitem[\protect\citeauthoryear{Van~Buuren}{Van~Buuren}{2007}]{vanBuuren.2007}
Van~Buuren, S. (2007).
\newblock {Multiple Imputation of Discrete and Continuous Data by Fully
  Conditional Specification}.
\newblock {\em Statistical Methods in Medical Research\/}~{\em 16\/}(3),
  219--242.

\bibitem[\protect\citeauthoryear{Van~Buuren}{Van~Buuren}{2018}]{vanBuuren.2018}
Van~Buuren, S. (2018).
\newblock {\em {Flexible Imputation of Missing Data}}.
\newblock Chapman \& Hall.

\bibitem[\protect\citeauthoryear{Van~Buuren, Brand, Groothuis-Oudshoorn, and
  Rubin}{Van~Buuren et~al.}{2006}]{vanBuuren.et.al.2006}
Van~Buuren, S., J.~P. Brand, C.~G. Groothuis-Oudshoorn, and D.~B. Rubin (2006).
\newblock {Fully Conditional Specification in Multivariate Imputation}.
\newblock {\em Journal of Statistical Computation and Simulation\/}~{\em
  76\/}(12), 1049--1064.

\bibitem[\protect\citeauthoryear{Van~Buuren and Groothuis-Oudshoorn}{Van~Buuren
  and Groothuis-Oudshoorn}{2011}]{vanBuuren.GroothuisOudshoorn.2011}
Van~Buuren, S. and K.~Groothuis-Oudshoorn (2011).
\newblock {{mice}: Multivariate Imputation by Chained Equations in R}.
\newblock {\em Journal of Statistical Software\/}~{\em 45\/}(3), 1--67.

\bibitem[\protect\citeauthoryear{Zubizarreta}{Zubizarreta}{2015}]{Zubizarreta.2015}
Zubizarreta, J.~R. (2015).
\newblock {Stable Weights That Balance Covariates for Estimation With
  Incomplete Outcome Data}.
\newblock {\em Journal of the American Statistical Association\/}~{\em
  110\/}(511), 910--922.

\end{thebibliography}

\newpage

\begin{center}
{\huge Supplementary Information}
\end{center}

\begin{appendices} 

\renewcommand\thetable{\thesection.\arabic{table}}    
\renewcommand\thefigure{\thesection.\arabic{figure}}    
\renewcommand\theequation{\thesection.\arabic{equation}}
\setcounter{figure}{0}    
\setcounter{table}{0}    
\setcounter{equation}{0}

\section{Additional Figures}

\begin{figure}[ht]
\centering
\includegraphics[width=\textwidth]{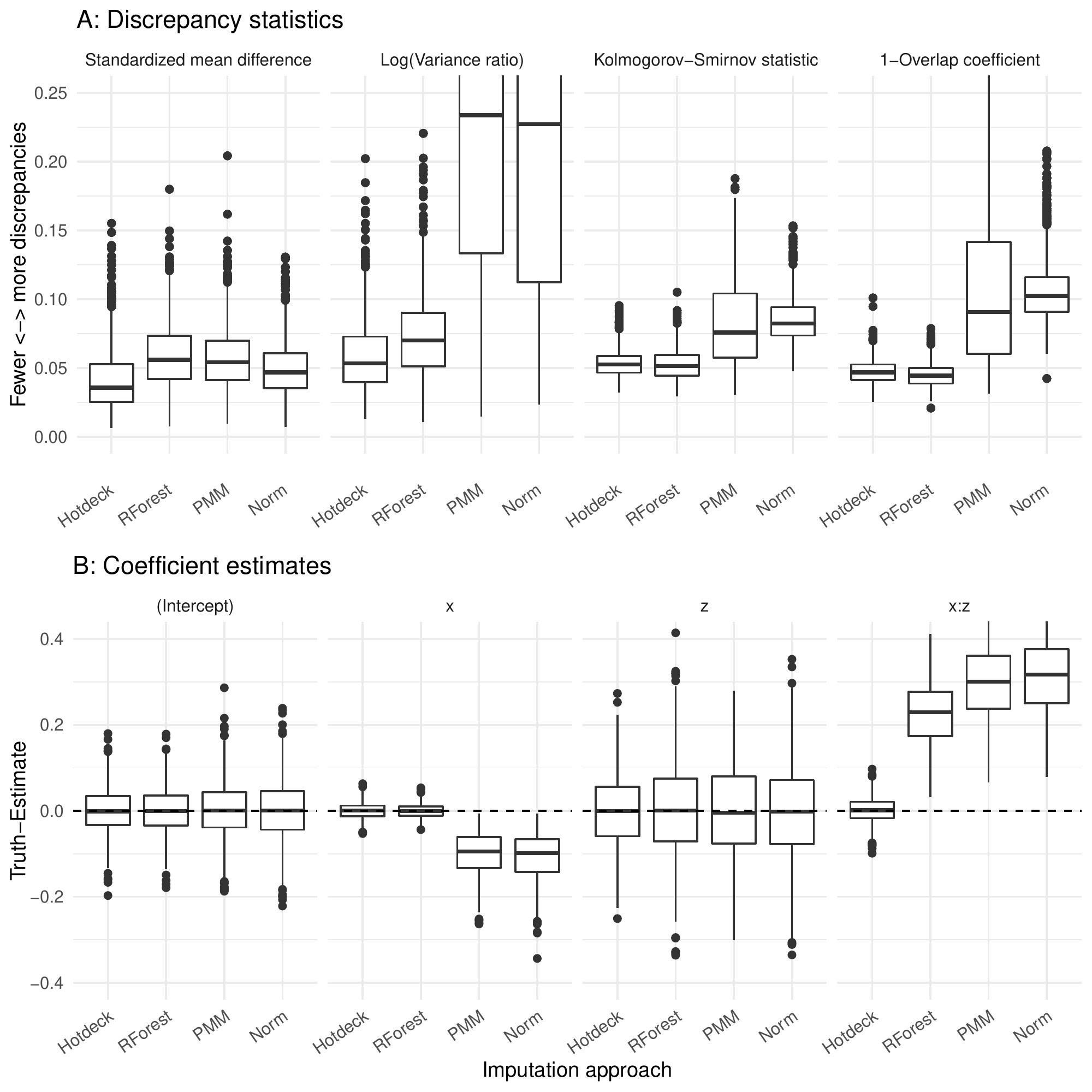}
\caption{Summary of 1,000 Monte Carlo simulations. Panel A: Box plots of four discrepancy statistics comparing the weighted density of observed and imputed values. \emph{Weights are constructed to balance the first and second moment of the covariate densities.} Panel B: Box plots of the bias in the (pooled) coefficient estimates of a linear model $y = x \cdot z + \epsilon$. Each dataset was imputed five times using random hot-deck imputation (Hotdeck), random forest imputation (RForest), predictive mean matching (PMM) and normal model imputation (Norm). Some box plots are clipped to increase readability. \label{mc_xx}}
\end{figure}

\begin{figure}[!ht]
\centering
\includegraphics[width=\textwidth]{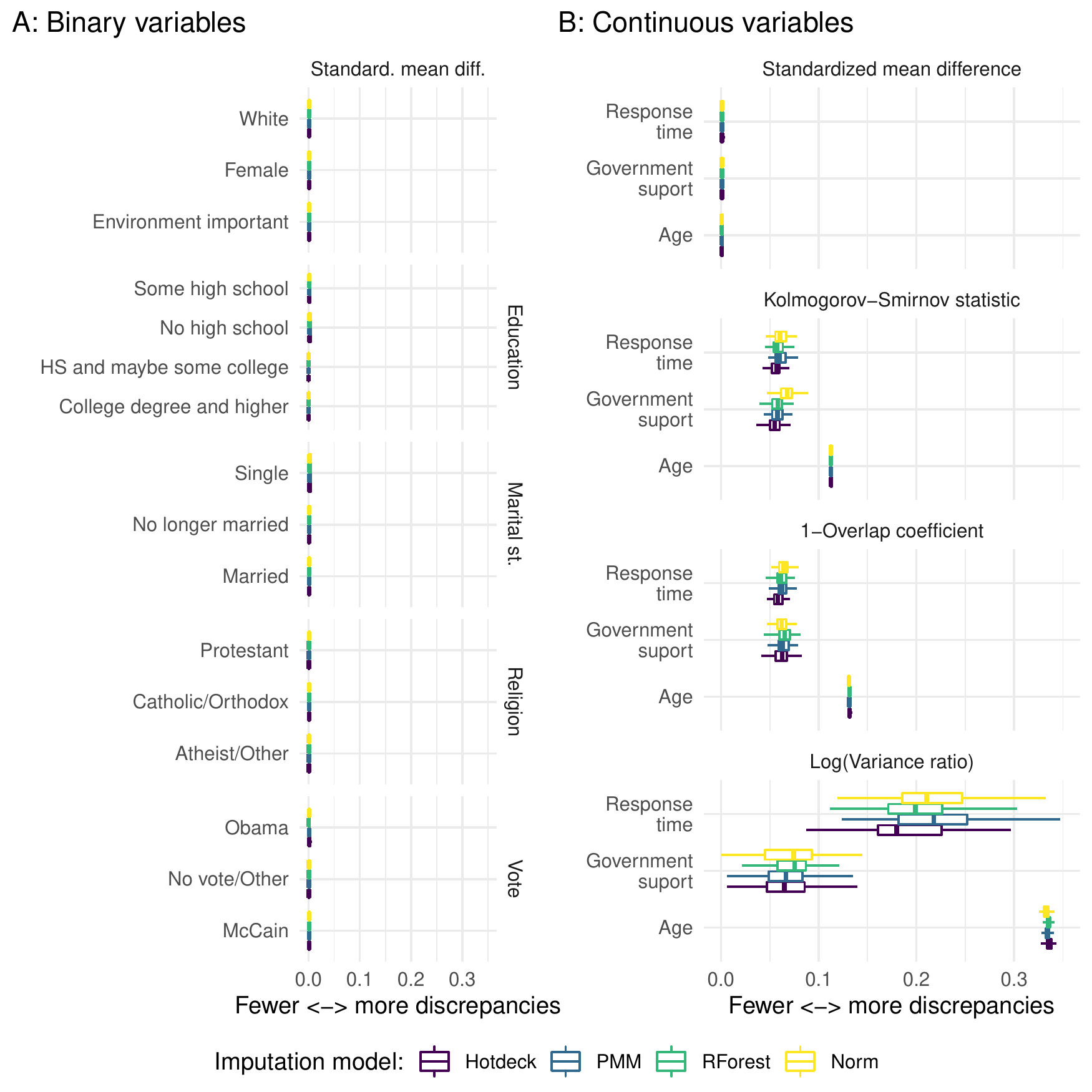}
\caption{Density of discrepancy statistics comparing observations with imputed and observed values of the variable \textbf{income} across 100 imputed datasets. Some outliers are removed, and some upper whiskers are clipped to increase readability. \label{fig_sm_income}}
\end{figure}

\begin{figure}[!ht]
\centering
\includegraphics[width=\textwidth]{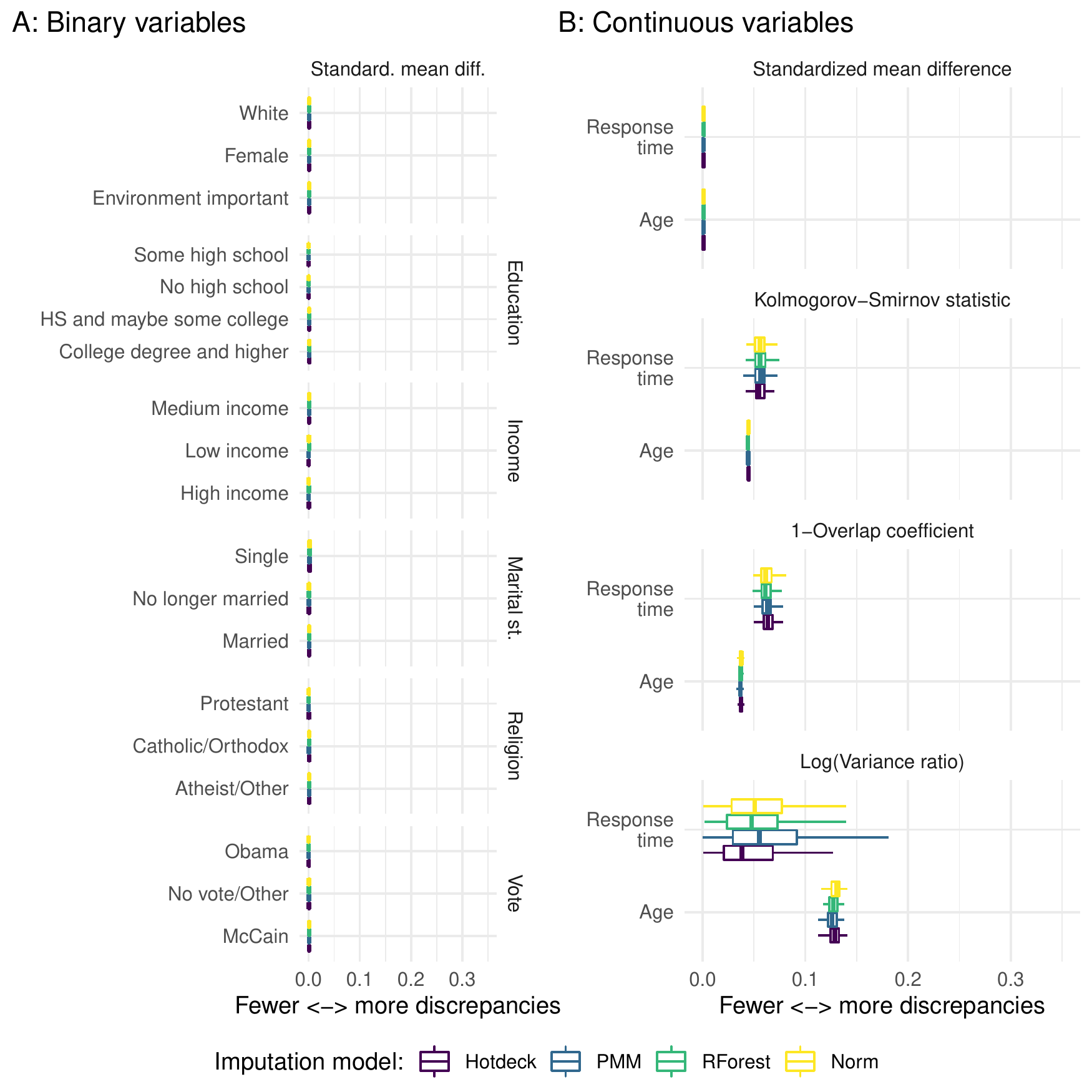}
\caption{Density of discrepancy statistics comparing observations with imputed and observed values of the variable \textbf{government support} across 100 imputed datasets. Some outliers are removed, and some upper whiskers are clipped to increase readability. \label{fig_sm_jobs_r}}
\end{figure}

\begin{figure}[!ht]
\centering
\includegraphics[width=\textwidth]{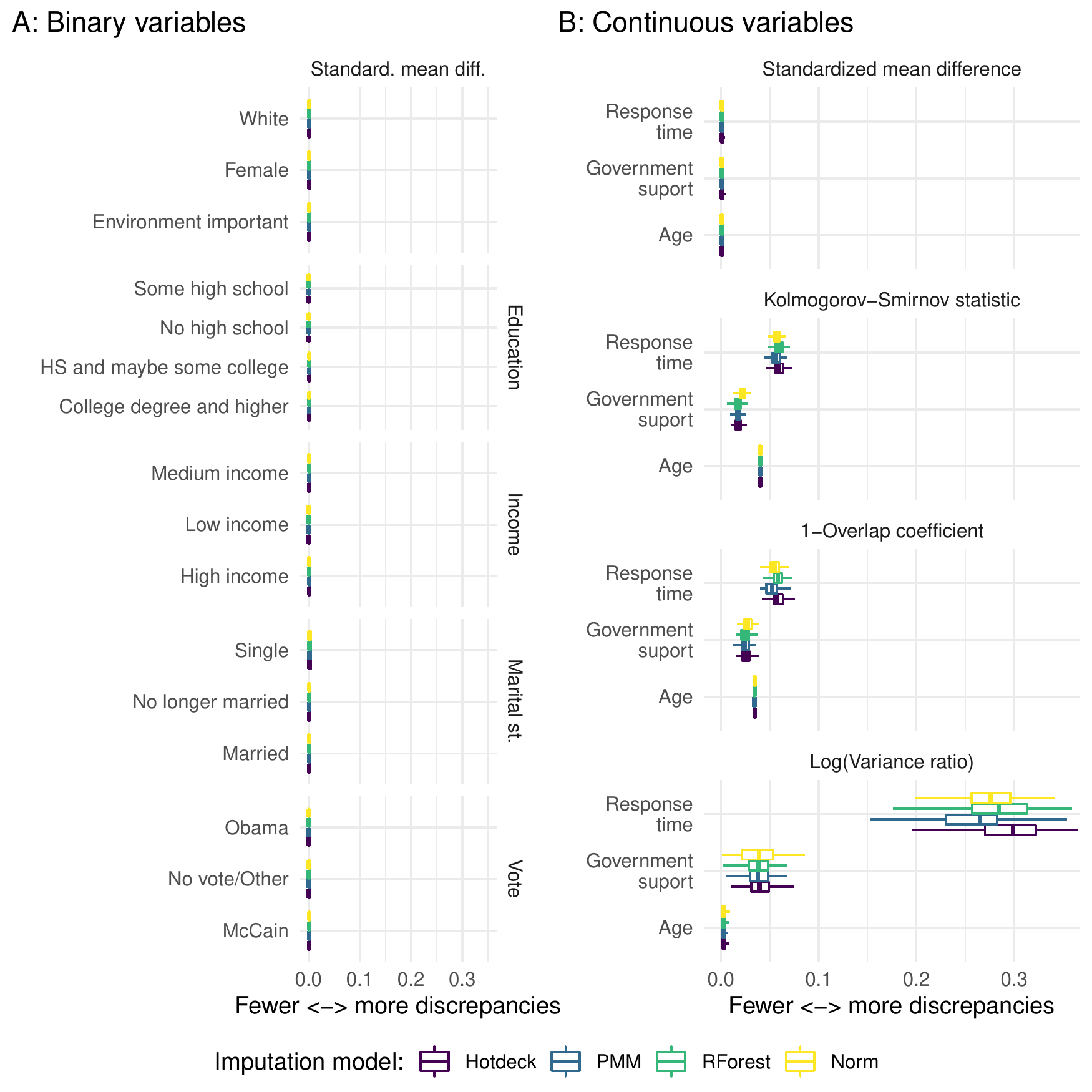}
\caption{Density of discrepancy statistics comparing observations with imputed and observed values of the variable \textbf{religion} across 100 imputed datasets. Some outliers are removed, and some upper whiskers are clipped to increase readability. \label{fig_sm_religion}}
\end{figure}

\begin{figure}[!ht]
\centering
\includegraphics[width=\textwidth]{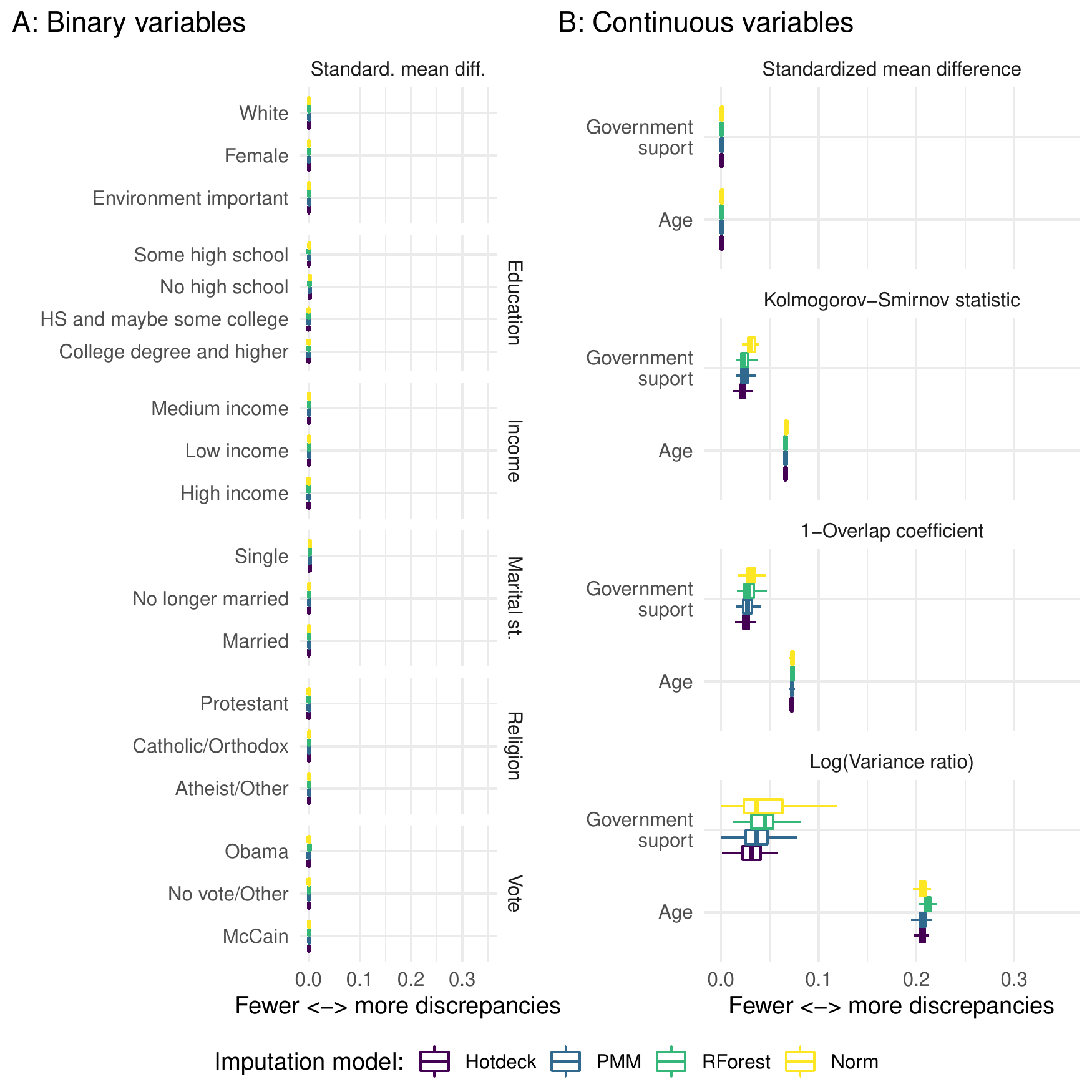}
\caption{Density of discrepancy statistics comparing observations with imputed and observed values of the variable \textbf{response time} across 100 imputed datasets. Some outliers are removed, and some upper whiskers are clipped to increase readability. \label{fig_sm_time}}
\end{figure}

\begin{figure}[!ht]
\centering
\includegraphics[width=\textwidth]{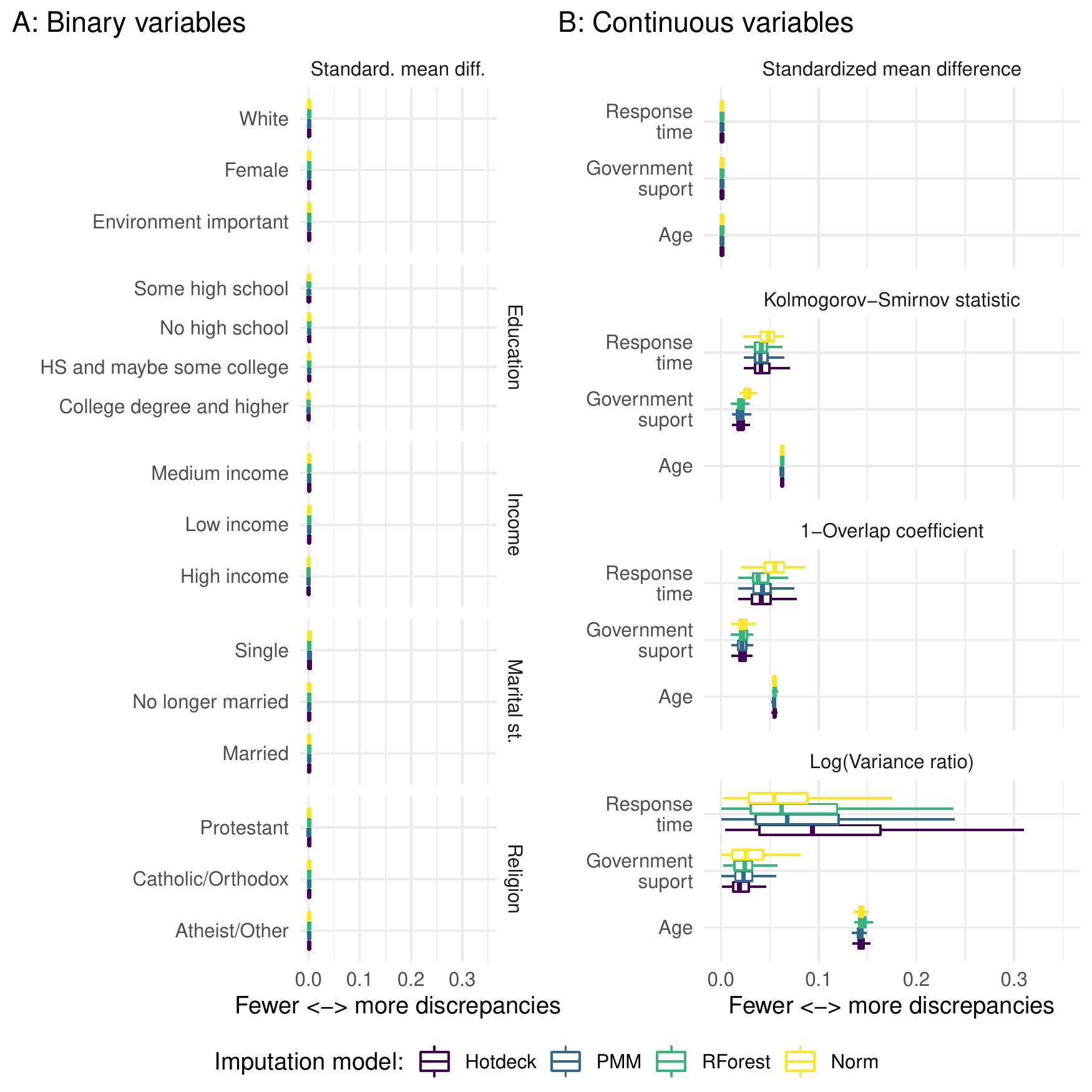}
\caption{Density of discrepancy statistics comparing observations with imputed and observed values of the variable \textbf{vote} across 100 imputed datasets. Some outliers are removed, and some upper whiskers are clipped to increase readability. \label{fig_sm_vote}}
\end{figure}

\begin{figure}[!ht]
\centering
\includegraphics[width=\textwidth]{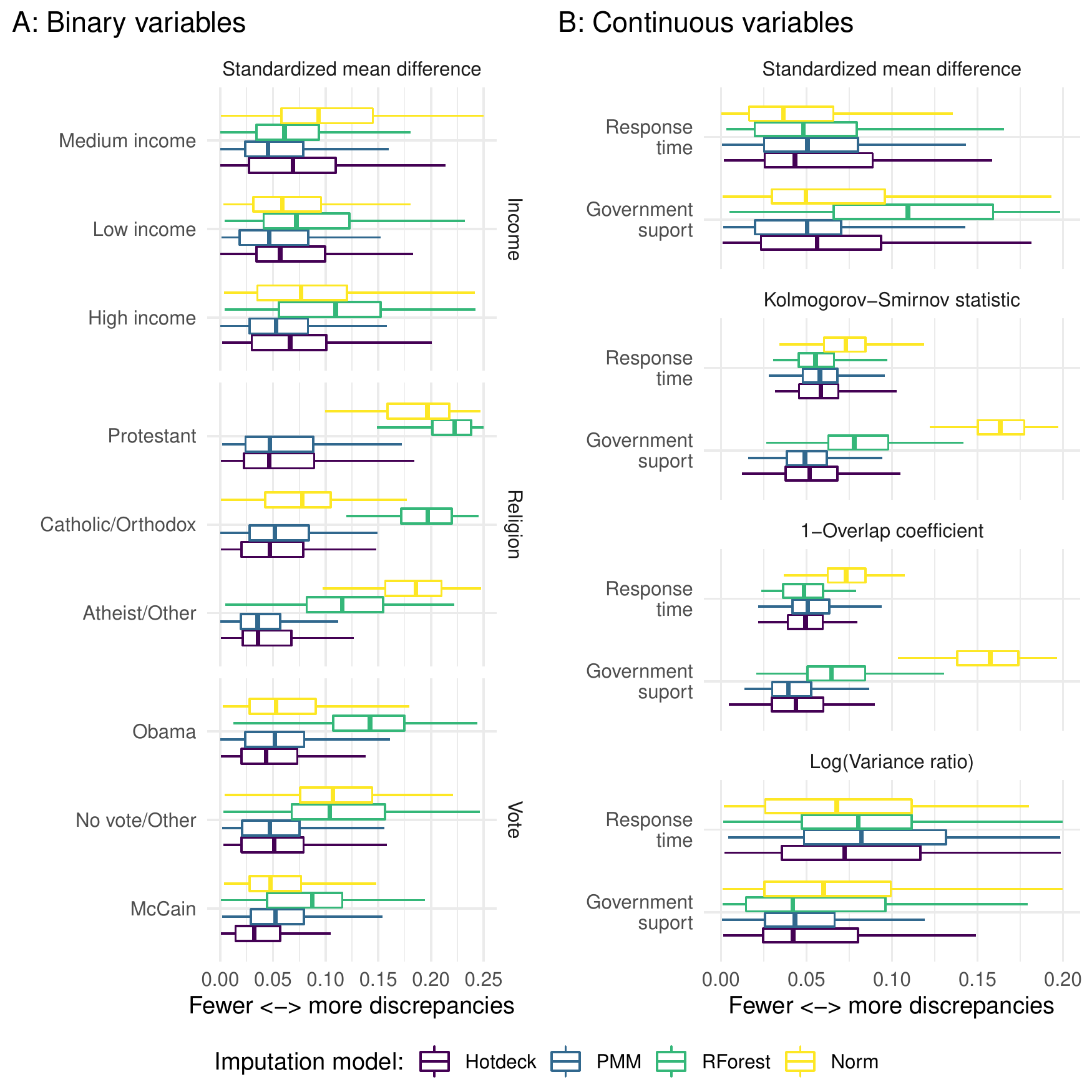}
\caption{\textbf{Additional balancing of interactions}. Box plots of discrepancy statistics comparing the weighted density of observed and imputed values across 100 imputed datasets. Some outliers are removed, and some upper whiskers are clipped to increase readability.\label{fig_int}}
\end{figure}

\end{appendices}

\end{document}